
%
%
%
%
\input amstex
\documentstyle{vanilla}
%
%
\def\myPS#1#2{#2}
%
%
%
\myPS{
\font\tenrm=cmr10 at 12truept
\font\teni=cmmi10 at 12truept
\font\tensy=cmsy10 at 12truept
\font\tenex=cmex10 at 12truept
\font\tenbf=cmbx10 at 12truept
\font\tensl=cmsl10 at 12truept
\font\tentt=cmtt10 at 12truept
\font\tenit=cmti10 at 12truept
\font\tenib=cmmib10 at 12truept
\font\tensb=cmbsy10 at 12truept
\font\ninebf=cmbx10 at 10.8truept
\font\ninerm=cmr10 at 10.8truept
\font\nineit=cmti10 at 10.8truept
 at 9.6truept
\font\eightrm=cmr7 at 9.6truept
\font\eightit=cmti7 at 9.6truept
\font\eighti=cmmi7 at 9.6truept
\font\eightsy=cmsy7 at 9.6truept
\font\sevenrm=cmr7 at 8.4truept
\font\seveni=cmmi7 at 8.4truept
\font\sevensy=cmsy7 at 8.4truept
\font\sevenbf=cmbx7 at 8.4truept
\font\fiverm=cmr5 at 6truept
\font\fivei=cmmi5 at 6truept
\font\fivesy=cmsy5 at 6truept
\font\fivebf=cmbx5 at 6truept}
{
\font\tenrm=cmr10 at 12truept
\font\teni=cmmi10 at 12truept
\font\tensy=cmsy10 at 12truept
\font\tenex=cmex10 at 12truept
\font\tenbf=cmbx10 at 12truept
\font\tensl=cmsl10 at 12truept
\font\tentt=cmtt10 at 12truept
\font\tenit=cmti10 at 12truept
\font\tenib=cmmib10 at 12truept
\font\tensb=cmbsy10 at 12truept
\font\ninebf=cmbx9 at 10.8truept
\font\ninerm=cmr9 at 10.8truept
\font\nineit=cmti9 at 10.8truept
 at 9.6truept
\font\eightrm=cmr8 at 9.6truept
\font\eightit=cmti8 at 9.6truept
\font\eighti=cmmi8 at 9.6truept
\font\eightsy=cmsy8 at 9.6truept
\font\sevenrm=cmr7 at 8.4truept
\font\seveni=cmmi7 at 8.4truept
\font\sevensy=cmsy7 at 8.4truept
\font\sevenbf=cmbx7 at 8.4truept
\font\fiverm=cmr5 at 6truept
\font\fivei=cmmi5 at 6truept
\font\fivesy=cmsy5 at 6truept
\font\fivebf=cmbx5 at 6truept}
\skewchar\teni='177 \skewchar\seveni='177 \skewchar\fivei='177
\skewchar\tensy='60 \skewchar\sevensy='60 \skewchar\fivesy='60
\textfont0=\tenrm \scriptfont0=\sevenrm \scriptscriptfont0=\fiverm
\def\rm{\fam0 \tenrm}
\textfont1=\teni  \scriptfont1=\seveni  \scriptscriptfont1=\fivei
 
\textfont2=\tensy \scriptfont2=\sevensy \scriptscriptfont2=\fivesy

\textfont3=\tenex \scriptfont3=\tenex   \scriptscriptfont3=\tenex
\newfam\itfam \def\it{\fam\itfam\tenit} \textfont\itfam=\tenit
\newfam\slfam \def\sl{\fam\slfam\tensl} \textfont\slfam=\tensl
\newfam\bffam \def\bf{\fam\bffam\tenbf} \textfont\bffam=\tenbf
  \scriptfont\bffam=\sevenbf   \scriptscriptfont\bffam=\fivebf
\newfam\ttfam  \textfont\ttfam=\tentt

\def\AMPTY{1}
\def\BPA{2}
\def\AP{3}
\def\Ba{4}
\def\FZ{5}
\def\BEa{6}
\def\BEb{7}
\def\BEc{8}
\def\BEd{9}
\def\Za{10}
\def\Sy{11}
\def\Ga{12}
\def\SU{13}
\def\ref#1{$^{#1}$}
\def\refp#1{.$^{#1}\,$}
\def\refc#1{,$^{#1}$}

\def\delete#1{}

\def\sectiontitle#1\par{\vskip0pt plus.1\vsize\penalty-250
 \vskip0pt plus-.1\vsize\bigskip\vskip\parskip
 \message{#1}\leftline{\tenbf#1}\nobreak\vglue 5pt}
\def\qed{\hbox{${\vcenter{\vbox{
    \hrule height 0.4pt\hbox{\vrule width 0.4pt height 6pt
    \kern5pt\vrule width 0.4pt}\hrule height 0.4pt}}}$}}
\TagsOnRight
\hsize=6.0truein
\vsize=8.5truein
\parindent=15pt
\nopagenumbers
\baselineskip=14pt
\def\wb{{\overline W}}

\def\wf{W^{(f)}}
\def\wbf{{\overline W}\vphantom{W}^{(f)}}
\def\wu{W\vphantom{\wbf}}
\def\tb{{\bar \theta}}
\def\pb{{\bar \phi}}

\def\half{{\textstyle\frac{1}{2}}}
\def\halfs{{\scriptstyle\frac{1}{2}}}
\def \up#1#2{\leavevmode \raise #1ex\hbox{#2}}

\baselineskip=14pt
\headline{\ifnum\pageno=1\hfil\else%
{\ifodd\pageno\rightheadline \else \leftheadline\fi}\fi}
\def\rightheadline{\eightit Helen Au-Yang and Jacques H.H.\ Perk%
\hfil---$\!$---\hfil
The N$\to\!\!\infty$ Limit of the Chiral Potts Model\quad\eightrm\folio}
\def\leftheadline{\eightrm\folio\quad
\eightit
Helen Au-Yang and Jacques H.H.\ Perk\hfil---$\!$---\hfil
The N$\to\!\!\infty$ Limit of the Chiral Potts Model}
\voffset=2\baselineskip
\strut\vskip -3\baselineskip hep-th/9305171\hfill OSU-SM-93-01
\vskip 2\baselineskip
\centerline{\tenbf
THE $\hbox{\tenib N}\,\hbox{\tensb \char'041}\,\hbox{\tensb \char'061}$
LIMIT OF THE CHIRAL POTTS MODEL
}
\vglue 24pt
\centerline{\eightrm HELEN AU-YANG and JACQUES~H.H.~PERK
\footnote"$^*$"{\eightrm \baselineskip=10pt
Supported in part by NSF Grants Nos.\ DMS 91-06521 and INT 91-12563}}
\centerline{\eightit Department of Physics, Oklahoma State University}
\centerline{\eightit Stillwater, OK 74078-0444, USA
\footnote"$^{\dagger}$"{\eightrm \baselineskip=10pt
Work done at
RIMS, Kyoto University, Sakyo-ku, Kyoto 606, Japan}}
\vglue 20pt
\baselineskip=12pt
\vglue 20pt
\centerline{\eightrm ABSTRACT}
{\rightskip=3pc
\leftskip=3pc
\eightrm\parindent=1pc
We consider the $\hbox{\eighti N}\,\hbox{\eightsy \char'041}\,
\hbox{\eightsy \char'061}$ limits of the $\hbox{\eighti N}$-state chiral Potts
model. We find new weights that satisfy the star-triangle relations with
spin variables either taking all the integer values or having values from a
continous interval. The models provide chiral generalizations of
Zamolodchikov's Fishnet Model.
\vglue12pt}
\baselineskip=14pt
\parindent=3pc
\tenrm
\sectiontitle{1. Introduction}

In the integrable $N$-state chiral Potts model\ref{\AMPTY} the Boltzmann
weights $W(a-b)$ and $\wb(a-b)$ --- corresponding to the horizontal and
vertical pair interactions between spins in states $a$ and $b$ --- satisfy
the ``star-triangle" equation\ref{\BPA,\AP}
$$
\sum^{N}_{d=1}\, \wb_{qr}(b-d)\, W_{pr}(a-d)\, \wb_{pq}(d-c)
= R_{pqr}\, W_{pq}(a-b)\, \wb_{pr}(b-c)\, W_{qr}(a-c).\eqno{(1)}
$$
The resulting weights have a product form\refp{\BPA,\AP} Here we
rewrite\ref{\Ba} these results as
$$
\eqalignno{
{W_{pq}(n)\over W_{pq}(0)} &= \Bigl({\sin \theta_p \sin \phi_q \over
\sin \theta_q \sin \phi_p }\Bigr)^{n/2N}
\prod^{n}_{j=1} {\sin [\pi(j-\half)/N -(\theta_q-\phi_p)/2N]
\over \sin[\pi(j-\half)/N +(\phi_q-\theta_p)/2N]},&{(2)}\cr
{\wb_{pq}(n)\over \wb_{pq}(0)} &= \Bigl({\sin \theta_p \sin \theta_q \over
\sin \phi_p \sin \phi_q }\Bigr)^{n/2N}
\prod^{n}_{j=1}
{\sin [\pi(j-1)/N +(\phi_q-\phi_p)/2N]
\over \sin[\pi j/N -(\theta_q-\theta_p)/2N]} ,&{(3)}
}
$$
using the substitutions
$$
{b_p\over c_p}=\omega^{\halfs}e^{i\theta_p/N},\quad
{a_p\over d_p}=e^{i\phi_p/N},\quad
\omega=e^{2\pi i/N},\quad
{c_p\over d_p}=
\left({e^{i\phi_p}\sin\phi_p\over e^{i\theta_p}\sin\theta_p}\right)^{1/2N},
\eqno{(4)}
$$
where $\theta_p$ and $\phi_p$ differ by a factor $N$ from those of
Baxter\refp{\Ba} The last equality in (4) follows from the integrability
conditions\ref{\BPA,\AP} or
$$
k={\sin\half(\theta_p-\phi_p)\over\sin\half(\theta_p+\phi_p)}=
{\sin\half(\theta_q-\phi_q)\over\sin\half(\theta_q+\phi_q)}.\eqno{(5)}
$$
The Fourier transforms, which are the weights after duality
transformation\refc{\BPA,\AP}
$$
\wf_{pq}(n)=N^{-1}\sum_{j=0}^{N-1}\, \omega^{-jn}\, W_{pq}(j),\qquad
\wbf_{pq}(n))=N^{-1}\sum_{j=0}^{N-1}\, \omega^{-jn} \,\wb_{pq}(j),\eqno{(6)}
$$
can be rewritten as
$$
\eqalignno{
{\wf_{pq}(n)\over\wf_{pq}(0)} &=
{e^{i n(\theta_q-\phi_q+\theta_p-\phi_p)/2N}}
\prod^{n}_{j=1} {\sin [\pi( j-1)/N +(\tb_q-\tb_p)/2N]
\over \vphantom{1^{1^1}}\sin[\pi j/N -(\pb_q-\pb_p)/2N]},&{(7)}\cr
{\wbf_{pq}(n)\over \wbf_{pq}(0)} &=
{e^{i n(\theta_q-\phi_q-\theta_p+\phi_p)/2N}}
\prod^{n}_{j=1}
{\sin [\pi(j-\half)/N -(\tb_q-\pb_p)/2N]
\over \vphantom{1^{1^1}}\sin[\pi(j-\half)/N +(\pb_q-\tb_p)/2N]} ,&{(8)}
}
$$
where
$$
\pb_p=\half(\theta_p+\phi_p)+\half i\log{\sin\phi_p\over\sin\theta_p},\quad
\tb_p=\half(\theta_p+\phi_p)-\half
i\log{\sin\phi_p\over\sin\theta_p}.\eqno{(9)}
$$
By direct substitution we can show that if the weights satisfy the
star-triangle equation (1) then their Fourier transforms satisfy
the star-triangle equation
$$
{N\over R_{pqr}} \wbf_{qr}(a) \, \wf_{pr}(b) \, \wbf_{pq}(a+b)
=\sum^{N-1}_{d=0}\, \wf_{pq}(b-d)\, \wbf_{pr}(a+b-d)\,
\wf_{qr}(d).\eqno{(10)}
$$
\noindent This equation has the exact same form as equation (1), as can be seen
replacing $a \to a-b$, $b \to b-c$, $a+b \to a-c$, and $c+d \to d$.

\par For $\theta_p=\phi_p$, $\theta_q=\phi_q$ we recover the
self-dual Fateev and Zamolodchikov\ref{\FZ} solution
with\vphantom{$1^{1^1}$}
${\wf(n)/\wf(0)}={\wb(n)/\wb(0)}$, ${\wbf(n)/\wbf(0)}={\wu(n)/\wu(0)}$.

\sectiontitle{2. The $\hbox{\tenib N} \hbox{\tensb\char'041}
\hbox{\tensb\char'061}$ 
Limit of the Boltzmann Weights}

We shall now obtain the $N \to \infty$ limit of the Boltzmann weights
(2) and (3) or their dual weights (7) and (8). Note that these all have
the product form
$$
{W(n)\over W(0)} = A ^{n/N} \prod^{n}_{j=1}
 {\sin\left(\vphantom{1^1}\pi(j+\alpha-1)/N\right) \over
 \sin\left(\vphantom{1^1}\pi(j+\beta-1)/N\right)},\quad
A={\sin(\pi\beta)\over\sin(\pi\alpha)},\eqno{(11)}
$$
with $\alpha$ and $\beta$ given constants depending on parameters
$\theta_p$, $\theta_q$, $\phi_p$, and $\phi_q$ satisfying (5). Also,
the condition on $A$ guarantees that $W(n+N)=W(n)$. We can rewrite formula
(11) in terms of a convergent series in powers of $1/N$, i.e.
$$\eqalignno{
\log&{W(n)\over W(0)} = \log\left(A ^{n/N}
\prod^{n}_{j=1} {j+\alpha-1 \over j+\beta-1}\right)\cr
&+\sum_{l=0}^{\infty}{B_{l+1}(\alpha)-B_{l+1}(\beta)\over(l+1)!}
{\left({2\pi\over N}\right)}^l {\left({d\over dz}\right)}^l
{\left.{\log\left({\sin z\over z}\right)}\right\vert}_{z=\pi n/N},&{(12)}}
$$
where $0<n<N$ and we used the functional relations of Bernoulli
polynomials\ref{\BEa} $\sum_{m=0}^p \left({p\atop m}\right) B_m(x) y^{p-m} =
B_p(x+y)$ and $B_p(x+1)-B_p(x)=p x^{p-1}$.
\vskip 4pt\par
There are three regimes for the limit, the first having $N \to \infty$, while
$n$ remains finite. In this case, (12) results in
$$
\displaylines{\hbox{I.}\hfill{}
{W(n)\over W(0)} =
\prod^{n}_{j=1}\, {j+\alpha-1 \over j+\beta-1} =
\prod^{-n}_{j=1}\, {j-\beta \over j-\alpha} =
{\Gamma(n+\alpha)\Gamma(\beta)\over\Gamma(n+\beta)\Gamma(\alpha)},\quad
-\infty < n < \infty.\hfill{}{(13)}\cr}
$$

The second regime has $N$, $n \to \infty$ such that $2\pi n/N\to x$ remains
finite. Consequently, the weights $W(n)$ in (11), which originally took $N$
different values and which were periodic modulo $N$, now depend on the
continuous spin values $x$ and they are periodic modulo $2\pi$. Using
the asymptotic formula\ref{\BEb} for $\log\Gamma(x)$, and $B_1(x)=x-\halfs$
we find that (12) in this limit gives
$$
\displaylines{\hbox{II.}\hfill{}
W(x) = C\,
{A\vphantom{1^1}}^{\textstyle{x\over 2\pi}-\left[{x\over 2\pi}\right]}\,
{\left\vert\sin{\half x}\right\vert}^{\alpha-\beta},
\hfill{}{(14)}\cr}
$$
where $[x]$ stands for integral part of $x$ and
$$
C = W(0)\,{(N/\pi)}^{\alpha-\beta}\,\Gamma(\beta)\,/\Gamma(\alpha).\eqno{(15)}
$$
The results in Regimes I and II are consistent with duality transformation.
More precisely, if the limiting weights are in Regime II, their Fourier
transforms are in Regime I, and vice versa. Indeed, the infinite sum in the
Fourier transform can be summed using the formula\ref{\BEc}
$$
\sum_{n=-\infty}^{\infty} e^{inx} \prod_{j=1}^n {j+\alpha-1\over j+\beta-1}=
{2^{\beta-\alpha-1}\, \Gamma(1-\alpha)\Gamma(\beta)\over\Gamma(\beta-\alpha)}\,
\, e^{\halfs i(1-\alpha-\beta)(x-\pi)}
\,{\vert\sin\half x\vert}^{\beta-\alpha-1},\eqno{(16)}
$$
for $0<x<2\pi$ (and periodically extended) and with $\alpha$ and $\beta$ in
(13) calculated from (2) or (3), to give an identical formula for the weights
as in (14) for Regime II, with new values of $\alpha$ and $\beta$
corresponding to (7)-(9). This generalizes the large-$N$ limit of
Fateev and Zamolodchikov\refp{\FZ}

A third intermediate regime can also appear with $N$, $n \to \infty$ such
that $n/f(N)\to x$ for some function $f(N)$ that blows up slower than $N$.
We have
$$
\displaylines{\hbox{III.}\hfill{}
W(x) =
D\,{A\vphantom{1^1}}^{-\halfs{\rm sign}(x)}\,
{\left\vert x\right\vert}^{\alpha-\beta},\quad
-\infty < x < \infty,\hfill{}{(17)}\cr}
$$
which is a chiral generalization of Zamolodchikov's Fishnet Model\refp{\Za}
In (17), $D=C\,A^{\halfs}$, with $C$ given by (16). The sign function in (17)
emerges if we rewrite (11) for negative $n$ in the form (12), see also (13),
and compare constants in (17) for $x>0$ and $x<0$; then we can use
${\Gamma(1-\alpha)/\Gamma(1-\beta)}=A\,{\Gamma(\beta)/\Gamma(\alpha)}$, which
follows from $\Gamma(x)\Gamma(1-x)=\pi/\sin(\pi x)$, leading to an extra
factor $A$ for $x<0$. The sign function relates directly to the effect of the
integer part in (14) near $x=0$; in fact, by the identical reasoning for
Regime II, we need the same extra factor $A$ for $-2\pi<x<0$ in (14).

We note that in the previously known cases\ref{\FZ,\Za} $\alpha+\beta=1$ and
$A=1$. We have only one condition (11) on $A$ allowing the deformations
(13), (14), and (17) to provide integrable field theories with chirality.

\sectiontitle{3. The $\hbox{\tenib N} \hbox{\tensb\char'041}
\hbox{\tensb\char'061}$ 
Limit of the Chiral Potts Model}

Having obtained explicit prescriptions on how to take the $N \to \infty$ limit
of the Boltzmann weights of the $N$-state chiral Potts model, we are now in
a position to examine the various limits of the star-triangle equation (1).
The summation over $d$ must be split in several pieces as we must choose from
which regimes to take the three weights in the summand. As $N\to\infty$, we
can rigorously show that most pieces can be ignored.

The least complicated (but also a most interesting) case occurs when all the
$\beta-\alpha$ in the formulae of section 2 are between 0 and 1, defining a
principal domain for the spectral parameters. In this case, the three types
of large-$N$ behavior do not mix. More precisely, if we take the three spin
states $a$, $b$, and $c$ in (1) mutually separated according to one and the
same regime in the sense of the previous section --- so that we are taking
the same type of limit for all three weights in the right-hand side of (1) ---
the dominant part of the sum over spin state $d$ comes from the piece with
all three weights in the left-hand side of (1) from the same regime.
This is easily proved as the formulae in section 2 give full control over
the leading term and the corrections in the large-$N$ limit.

For example, taking Regime II, the star-triangle equations become
$$
\int^{2\pi}_0 dw \,\wb_{qr}(y-w)\, W_{pr}(x-w)\, \wb_{pq}(w-z)
= R\, W_{pq}(x-y)\, \wb_{pr}(y-z)\, W_{qr}(x-z),\eqno{(18)}
$$
after a suitable renormalization of $R$. This has the solution
$$
\eqalignno{
W_{pq}(x)&=C_{pq}\,{e\vphantom{1^1}}^{(\gamma_p-\gamma_q)
({\textstyle{x\over 2\pi}-\left[{x\over 2\pi}\right]})}\,
{\left\vert\sin{\half x}\right\vert}^{\lambda_p-\lambda_q},\cr
\wb_{pq}(x)&={\overline C}_{pq}\,{e\vphantom{1^1}}^{(\gamma_p+\gamma_q)
({\textstyle{x\over 2\pi}-\left[{x\over 2\pi}\right]})}\,
{\left\vert\sin{\half x}\right\vert}^{\lambda_q-\lambda_p-1},
&(19)}
$$
with
$$
\lambda_p={1\over\pi}\,\arctan\left({\sin\theta_p\over\cos\theta_p+k}\right),
\quad
\gamma_p=\half\log\left({1+2k\cos\theta_p+k^2\over1-k^2}\right),
\eqno{(20)}
$$
and similar formulae for $\lambda_q$, $\gamma_q$, $\lambda_r$, and $\gamma_r$,
as follows from the prescriptions given in the previous two sections. If
$\lambda_p<\lambda_q<\lambda_r<1+\lambda_p$ all six Boltzmann weights in (18)
are real and positive and the parameters are in the principal domain.

We remark that it is straightforward to show that the weights obtained
by just dropping the integral part $[x/2\pi]$ in (19) also
satisfy the same star-triangle equation (18). This solution can be viewed as
the Fateev-Zamolodchikov solution\ref{\FZ} with a site-dependent gauge
transformation. We emphasize that we have both numerically verified and
analytically proved that the ``chiral" weights (19) also satisfy (18).

Since $W_{pq}(x)$ and $\wb_{pq}(x)$ are now functions of $x$ modulo $2\pi$,
their Fourier transforms
$$
\wf_{pq}(j)={1 \over 2\pi} \int_{0}^{2\pi} dx \, e^{-ijx}\, W_{pq}(x),
\qquad
\wbf_{pq}(j)={1 \over 2\pi}\int_{0}^{2\pi} dx \, e^{-ijx} \,\wb_{pq}(x),
\eqno{(21)}
$$
are over all the integer value $j$, ranging from $-\infty$ to $\infty$.
Substituting (21) into (18), we find that the Fourier transforms satisfy
different star-triangle equations,
$$
{2\pi\over R}\, \wbf_{qr}(a) \, \wf_{pr}(b) \, \wbf_{pq}(a+b)
=\sum^{\infty}_{d=-\infty}\, \wf_{pq}(b-d)\, \wbf_{pr}(a+b-d)\,
\wf_{qr}(d),\eqno{(22)}
$$
in which the sum is over all integer values of $d$.

Taking Regime I instead of Regime II, leads to the proof of a star-triangle
equation of the form (21) instead of (18). This is an interesting identity
in the theory of generalized hypergeometric series generalizing the
Dougall-Ramanujan identity\refp{\BEd}

Choosing Regime III, we can prove a star-triangle equation of the
form (18), but with integration over $(-\infty,+\infty)$. This generalizes
Symanzik's integral\refc{\Sy} which Zamolodchikov used to prove the
star-triangle equation for the Fishnet Model\ref{\Za} and which further
provided the proof for the Fateev-Zamolodchikov model\ref{\FZ} via a conformal
transformation, $\xi=\tan \half w$.

Finally, as in our joint work with Baxter\refc{\BPA} we can let the $R$-matrix
be the product of four weights of any of the above types I, II, or III, i.e.
$$
R(a,b,c,d)=\wb_{p_1q_1}(a-c)\, W_{p_1q_2}(c-b)\,
\wb_{p_2q_2}(d-b) \,W_{p_2q_1}(a-d).\eqno{(23)}
$$
Then any such infinite-dimensional $R$-matrix satisfies the Yang-Baxter
equation. But our solutions are very different from those of Gaudin\refc{\Ga}
Shibukawa and Ueno\refp{\SU}

\def\references{\vglue12pt plus.1\vsize\penalty-250
 \vskip0pt plus-.1\vsize\bigskip\vskip\parskip
 \message{References}
 \leftline{\tenbf References}\nobreak\vglue 5pt
 \baselineskip=11pt\par}
\def\no#1#2\par{\item{#1.}#2\par}
\def\jr#1{{\nineit#1}}
\def\book#1{{\nineit#1}}
\def\vl#1{{\ninebf#1}}
\references\ninerm
\baselineskip=13pt 
\frenchspacing

\no {\AMPTY}
H. Au-Yang, B. M. McCoy, J. H. H. Perk, S. Tang, and M.-L. Yan,
\jr {Phys. Lett.}
\vl {123A}
(1987) 219.

\no {\BPA}
R. J. Baxter, J. H. H. Perk, and H. Au-Yang,
\jr {Phys. Lett.}
\vl {128A}
(1988) 138.

\no {\AP}
H. Au-Yang and J. H. H. Perk,
\jr {Adv. Stud. Pure Math.}
\vl {19}
(1989) 57.

\no {\Ba}
R. J. Baxter,
\jr {J. Stat. Phys.}
\vl {52}
(1988) 639.

\no {\FZ}
V. A. Fateev and A. B. Zamolodchikov,
\jr {Phys. Lett.}
\vl {92A}
(1982) 37, see also 35.

\no {\BEa}
\book {Higher Transcendental Functions}
Vol. 1,
\book {H. Bateman Manuscript Project},
eds. A. Erd\'elyi et al.,
(McGraw-Hill, New York, 1953),
eqs. 1.13 (2), (6).

\no {\BEb}
{Ref. \BEa},
eq. 1.18 (1).

\no {\BEc}
{Ref. \BEa},
eqs. 2.1.2 (6), 2.9 (1), (13), (22), (27).

\no {\BEd}
{Ref. \BEa},
eq. 4.5 (7).

\no {\Za}
A. B. Zamolodchikov,
\jr {Phys. Lett.}
\vl {97B}
(1980) 63.

\no {\Sy}
K. Symanzik,
\jr {Lett. Nuovo Cim.}
\vl {3}
(1972) 734.

\no {\Ga}
M. Gaudin,
\jr {J. Phys. France}
\vl {49}
(1988) 1857.

\no {\SU}
Y. Shibukawa and K. Ueno,
\jr {Lett. Math. Phys.}
\vl {25}
(1992) 239;
\jr {Waseda University preprint}.

\vfill\eject
\baselineskip=14pt
\parindent=3pc
\tenrm
\sectiontitle{Errata}

Formula (12) has the higher orders misprinted and is only correct to the
order needed in the actual $N\to\infty$ limits presented in this work. The
result correct in higher order can be found in (3.15) with (3.13) of
math.QA/9906029 which has been published as Physica A {\bf 268} (1999)
175--206. That work goes also into far more detail in the subject matter.

\end
%
\catcode`\@=11
\font\tensmc=cmcsc10      
\def\smc{\tensmc}

\def\hcorrection#1{\advance\hoffset by #1 }
\def\vcorrection#1{\advance\voffset by #1 }
\def\wlog#1{}
\newif\iftitle@
\outer\def\title{\title@true\vglue 24\p@ plus 12\p@ minus 12\p@
   \bgroup\let\\=\cr\tabskip\centering
   \halign to \hsize\bgroup\tenbf\hfill\ignorespaces##\unskip\hfill\cr}
\def\endtitle{\cr\egroup\egroup\vglue 18\p@ plus 12\p@ minus 6\p@}
\outer\def\author{\iftitle@\vglue -18\p@ plus -12\p@ minus -6\p@\fi\vglue
    12\p@ plus 6\p@ minus 3\p@\bgroup\let\\=\cr\tabskip\centering
    \halign to \hsize\bgroup\smc\hfill\ignorespaces##\unskip\hfill\cr}
\def\endauthor{\cr\egroup\egroup\vglue 18\p@ plus 12\p@ minus 6\p@}
\outer\def\heading{\bigbreak\bgroup\let\\=\cr\tabskip\centering
    \halign to \hsize\bgroup\smc\hfill\ignorespaces##\unskip\hfill\cr}
\def\endheading{\cr\egroup\egroup\nobreak\medskip}

\outer\def\endproclaim{\par\ifdim\lastskip<\medskipamount\removelastskip
  \penalty 55 \fi\medskip\rm}
\outer\def\demo#1{\par\ifdim\lastskip<\smallskipamount\removelastskip
    \smallskip\fi\noindent{\smc\ignorespaces#1\unskip:\enspace}\rm
      \ignorespaces}

\newcount\footmarkcount@
\footmarkcount@=1
\def\makefootnote@#1#2{\insert\footins{\interlinepenalty=100
  \splittopskip=\ht\strutbox \splitmaxdepth=\dp\strutbox
  \floatingpenalty=\@MM
  \leftskip=\z@\rightskip=\z@\spaceskip=\z@\xspaceskip=\z@
  \noindent{#1}\footstrut\rm\ignorespaces #2\strut}}
\def\footnote{\let\@sf=\empty\ifhmode\edef\@sf{\spacefactor
   =\the\spacefactor}\/\fi\futurelet\next\footnote@}
\def\footnote@{\ifx"\next\let\next\footnote@@\else
    \let\next\footnote@@@\fi\next}
\def\footnote@@"#1"#2{#1\@sf\relax\makefootnote@{#1}{#2}}
\def\footnote@@@#1{$^{\number\footmarkcount@}$\makefootnote@
   {$^{\number\footmarkcount@}$}{#1}\global\advance\footmarkcount@ by 1 }

\hyphenation{man-u-script man-u-scripts ap-pen-dix ap-pen-di-ces}
\hyphenation{data-base data-bases}
\ifx\amstexloaded@\relax\catcode`\@=13
  \endinput\else\let\amstexloaded@=\relax\fi
\newlinechar=`\^^J
\def\eat@#1{}
\def\Space@.{\futurelet\Space@\relax}
\Space@. %
\newhelp\athelp@
{Only certain combinations beginning with @ make sense to me.^^J
Perhaps you wanted \string\@\space for a printed @?^^J
I've ignored the character or group after @.}
\def\futureletnextat@{\futurelet\next\at@}
{\catcode`\@=\active
\lccode`\Z=`\@ \lowercase
{\gdef@{\expandafter\csname futureletnextatZ\endcsname}
\expandafter\gdef\csname atZ\endcsname
   {\ifcat\noexpand\next a\def\next{\csname atZZ\endcsname}\else
   \ifcat\noexpand\next0\def\next{\csname atZZ\endcsname}\else
    \def\next{\csname atZZZ\endcsname}\fi\fi\next}
\expandafter\gdef\csname atZZ\endcsname#1{\expandafter
   \ifx\csname #1Zat\endcsname\relax\def\next
     {\errhelp\expandafter=\csname athelpZ\endcsname
      \errmessage{Invalid use of \string@}}\else
       \def\next{\csname #1Zat\endcsname}\fi\next}
\expandafter\gdef\csname atZZZ\endcsname#1{\errhelp
    \expandafter=\csname athelpZ\endcsname
      \errmessage{Invalid use of \string@}}}}
\def\atdef@#1{\expandafter\def\csname #1@at\endcsname}
\newhelp\defahelp@{If you typed \string\define\space cs instead of
\string\define\string\cs\space^^J
I've substituted an inaccessible control sequence so that your^^J
definition will be completed without mixing me up too badly.^^J
If you typed \string\define{\string\cs} the inaccessible control sequence^^J
was defined to be \string\cs, and the rest of your^^J
definition appears as input.}
\newhelp\defbhelp@{I've ignored your definition, because it might^^J
conflict with other uses that are important to me.}
\def\define{\futurelet\next\define@}
\def\define@{\ifcat\noexpand\next\relax
  \def\next{\define@@}%
  \else\errhelp=\defahelp@
  \errmessage{\string\define\space must be followed by a control
     sequence}\def\next{\def\garbage@}\fi\next}
\def\undefined@{}
\def\preloaded@{}
\def\define@@#1{\ifx#1\relax\errhelp=\defbhelp@
   \errmessage{\string#1\space is already defined}\def\next{\def\garbage@}%
   \else\expandafter\ifx\csname\expandafter\eat@\string
         #1@\endcsname\undefined@\errhelp=\defbhelp@
   \errmessage{\string#1\space can't be defined}\def\next{\def\garbage@}%
   \else\expandafter\ifx\csname\expandafter\eat@\string#1\endcsname\relax
     \def\next{\def#1}\else\errhelp=\defbhelp@
     \errmessage{\string#1\space is already defined}\def\next{\def\garbage@}%
      \fi\fi\fi\next}
\def\famzero{\fam\z@}

\def\arctan{\mathop{\famzero arctan}\nolimits}

\def\cos{\mathop{\famzero cos}\nolimits}

\def\lim{\mathop{\famzero lim}}

\def\log{\mathop{\famzero log}\nolimits}

\def\sin{\mathop{\famzero sin}\nolimits}

\def\tan{\mathop{\famzero tan}\nolimits}

\def\textfont@#1#2{\def#1{\relax\ifmmode
    \errmessage{Use \string#1\space only in text}\else#2\fi}}
\textfont@\rm\tenrm
\textfont@\it\tenit
\textfont@\sl\tensl
\textfont@\bf\tenbf
\textfont@\smc\tensmc
\let\ic@=\/
\def\/{\unskip\ic@}
\def\textfonti{\the\textfont1 }
\def\t#1#2{{\edef\next{\the\font}\textfonti\accent"7F \next#1#2}}
\let\B=\=
\let\D=\.
\def~{\unskip\nobreak\ \ignorespaces}
{\catcode`\@=\active
\gdef\@{\char'100 }}
\atdef@-{\leavevmode\futurelet\next\athyph@}
\def\athyph@{\ifx\next-\let\next=\athyph@@
  \else\let\next=\athyph@@@\fi\next}
\def\athyph@@@{\hbox{-}}
\def\athyph@@#1{\futurelet\next\athyph@@@@}
\def\athyph@@@@{\if\next-\def\next##1{\hbox{---}}\else
    \def\next{\hbox{--}}\fi\next}
\def\.{.\spacefactor=\@m}
\atdef@.{\null.}
\atdef@,{\null,}
\atdef@;{\null;}
\atdef@:{\null:}
\atdef@?{\null?}
\atdef@!{\null!}
\def\srdr@{\thinspace}
\def\drsr@{\kern.02778em}
\def\sldl@{\kern.02778em}
\def\dlsl@{\thinspace}
\atdef@"{\unskip\futurelet\next\atqq@}
\def\atqq@{\ifx\next\Space@\def\next. {\atqq@@}\else
         \def\next.{\atqq@@}\fi\next.}
\def\atqq@@{\futurelet\next\atqq@@@}
\def\atqq@@@{\ifx\next`\def\next`{\atqql@}\else\def\next'{\atqqr@}\fi\next}
\def\atqql@{\futurelet\next\atqql@@}
\def\atqql@@{\ifx\next`\def\next`{\sldl@``}\else\def\next{\dlsl@`}\fi\next}
\def\atqqr@{\futurelet\next\atqqr@@}
\def\atqqr@@{\ifx\next'\def\next'{\srdr@''}\else\def\next{\drsr@'}\fi\next}

\def\textfontii{\the\textfont2 }
\def\{{\relax\ifmmode\lbrace\else
    {\textfontii f}\spacefactor=\@m\fi}
\def\}{\relax\ifmmode\rbrace\else
    \let\@sf=\empty\ifhmode\edef\@sf{\spacefactor=\the\spacefactor}\fi
      {\textfontii g}\@sf\relax\fi}
\def\nonhmodeerr@#1{\errmessage
     {\string#1\space allowed only within text}}
\def\linebreak{\relax\ifhmode\unskip\break\else
    \nonhmodeerr@\linebreak\fi}
\def\allowlinebreak{\relax
   \ifhmode\allowbreak\else\nonhmodeerr@\allowlinebreak\fi}
\newskip\saveskip@
\def\nolinebreak{\relax\ifhmode\saveskip@=\lastskip\unskip
  \nobreak\ifdim\saveskip@>\z@\hskip\saveskip@\fi
   \else\nonhmodeerr@\nolinebreak\fi}
\def\newline{\relax\ifhmode\null\hfil\break
    \else\nonhmodeerr@\newline\fi}
\def\nonmathaerr@#1{\errmessage
     {\string#1\space is not allowed in display math mode}}
\def\nonmathberr@#1{\errmessage{\string#1\space is allowed only in math mode}}
\def\mathbreak{\relax\ifmmode\ifinner\break\else
   \nonmathaerr@\mathbreak\fi\else\nonmathberr@\mathbreak\fi}
\def\nomathbreak{\relax\ifmmode\ifinner\nobreak\else
    \nonmathaerr@\nomathbreak\fi\else\nonmathberr@\nomathbreak\fi}
\def\allowmathbreak{\relax\ifmmode\ifinner\allowbreak\else
     \nonmathaerr@\allowmathbreak\fi\else\nonmathberr@\allowmathbreak\fi}
\def\pagebreak{\relax\ifmmode
   \ifinner\errmessage{\string\pagebreak\space
     not allowed in non-display math mode}\else\postdisplaypenalty-\@M\fi
   \else\ifvmode\penalty-\@M\else\edef\spacefactor@
       {\spacefactor=\the\spacefactor}\vadjust{\penalty-\@M}\spacefactor@
        \relax\fi\fi}
\def\nopagebreak{\relax\ifmmode
     \ifinner\errmessage{\string\nopagebreak\space
    not allowed in non-display math mode}\else\postdisplaypenalty\@M\fi
    \else\ifvmode\nobreak\else\edef\spacefactor@
        {\spacefactor=\the\spacefactor}\vadjust{\penalty\@M}\spacefactor@
         \relax\fi\fi}
\def\newpage{\relax\ifvmode\vfill\penalty-\@M\else\nonvmodeerr@\newpage\fi}
\def\nonvmodeerr@#1{\errmessage
    {\string#1\space is allowed only between paragraphs}}
\def\smallpagebreak{\relax\ifvmode\smallbreak
      \else\nonvmodeerr@\smallpagebreak\fi}
\def\medpagebreak{\relax\ifvmode\medbreak
       \else\nonvmodeerr@\medpagebreak\fi}
\def\bigpagebreak{\relax\ifvmode\bigbreak
      \else\nonvmodeerr@\bigpagebreak\fi}
\newdimen\captionwidth@
\captionwidth@=\hsize
\advance\captionwidth@ by -1.5in
\def\caption#1{}
\def\topspace#1{\gdef\thespace@{#1}\ifvmode\def\next
    {\futurelet\next\topspace@}\else\def\next{\nonvmodeerr@\topspace}\fi\next}
\def\topspace@{\ifx\next\Space@\def\next. {\futurelet\next\topspace@@}\else
     \def\next.{\futurelet\next\topspace@@}\fi\next.}
\def\topspace@@{\ifx\next\caption\let\next\topspace@@@\else
    \let\next\topspace@@@@\fi\next}
 \def\topspace@@@@{\topinsert\vbox to
       \thespace@{}\endinsert}
\def\topspace@@@\caption#1{\topinsert\vbox to
    \thespace@{}\nobreak
      \smallskip
    \setbox\z@=\hbox{\noindent\ignorespaces#1\unskip}%
   \ifdim\wd\z@>\captionwidth@
   \centerline{\vbox{\hsize=\captionwidth@\noindent\ignorespaces#1\unskip}}%
   \else\centerline{\box\z@}\fi\endinsert}
\def\midspace#1{\gdef\thespace@{#1}\ifvmode\def\next
    {\futurelet\next\midspace@}\else\def\next{\nonvmodeerr@\midspace}\fi\next}
\def\midspace@{\ifx\next\Space@\def\next. {\futurelet\next\midspace@@}\else
     \def\next.{\futurelet\next\midspace@@}\fi\next.}
\def\midspace@@{\ifx\next\caption\let\next\midspace@@@\else
    \let\next\midspace@@@@\fi\next}
 \def\midspace@@@@{\midinsert\vbox to
       \thespace@{}\endinsert}
\def\midspace@@@\caption#1{\midinsert\vbox to
    \thespace@{}\nobreak
      \smallskip
      \setbox\z@=\hbox{\noindent\ignorespaces#1\unskip}%
      \ifdim\wd\z@>\captionwidth@
    \centerline{\vbox{\hsize=\captionwidth@\noindent\ignorespaces#1\unskip}}%
    \else\centerline{\box\z@}\fi\endinsert}
\mathchardef\prime@="0230
\def\prime{{{}\prime@{}}}
\def\prim@s{\prime@\futurelet\next\pr@m@s}

\def\,{\relax\ifmmode\mskip\thinmuskip\else\thinspace\fi}
\def\!{\relax\ifmmode\mskip-\thinmuskip\else\negthinspace\fi}
\def\frac#1#2{{#1\over#2}}

\def\:{\nobreak\hskip.1111em{:}\hskip.3333em plus .0555em\relax}
\def\intic@{\mathchoice{\hskip5\p@}{\hskip4\p@}{\hskip4\p@}{\hskip4\p@}}
\def\negintic@
 {\mathchoice{\hskip-5\p@}{\hskip-4\p@}{\hskip-4\p@}{\hskip-4\p@}}
\def\intkern@{\mathchoice{\!\!\!}{\!\!}{\!\!}{\!\!}}
\def\intdots@{\mathchoice{\cdots}{{\cdotp}\mkern1.5mu
    {\cdotp}\mkern1.5mu{\cdotp}}{{\cdotp}\mkern1mu{\cdotp}\mkern1mu
      {\cdotp}}{{\cdotp}\mkern1mu{\cdotp}\mkern1mu{\cdotp}}}
\newcount\intno@
\def\iint{\intno@=\tw@\futurelet\next\ints@}
\def\iiint{\intno@=\thr@@\futurelet\next\ints@}
\def\iiiint{\intno@=4 \futurelet\next\ints@}
\def\idotsint{\intno@=\z@\futurelet\next\ints@}
\def\ints@{\findlimits@\ints@@}
\newif\iflimtoken@
\newif\iflimits@
\def\findlimits@{\limtoken@false\limits@false\ifx\next\limits
 \limtoken@true\limits@true\else\ifx\next\nolimits\limtoken@true\limits@false
    \fi\fi}
\def\multintlimits@{\intop\ifnum\intno@=\z@\intdots@
  \else\intkern@\fi
    \ifnum\intno@>\tw@\intop\intkern@\fi
     \ifnum\intno@>\thr@@\intop\intkern@\fi\intop}
\def\multint@{\int\ifnum\intno@=\z@\intdots@\else\intkern@\fi
   \ifnum\intno@>\tw@\int\intkern@\fi
    \ifnum\intno@>\thr@@\int\intkern@\fi\int}
\def\ints@@{\iflimtoken@\def\ints@@@{\iflimits@
   \negintic@\mathop{\intic@\multintlimits@}\limits\else
    \multint@\nolimits\fi\eat@}\else
     \def\ints@@@{\multint@\nolimits}\fi\ints@@@}
\def\Sb{_\bgroup\vspace@
        \baselineskip=\fontdimen10 \scriptfont\tw@
        \advance\baselineskip by \fontdimen12 \scriptfont\tw@
        \lineskip=\thr@@\fontdimen8 \scriptfont\thr@@
        \lineskiplimit=\thr@@\fontdimen8 \scriptfont\thr@@
        \Let@\vbox\bgroup\halign\bgroup \hfil$\scriptstyle
            {##}$\hfil\cr}
\def\endSb{\crcr\egroup\egroup\egroup}
\def\Sp{^\bgroup\vspace@
        \baselineskip=\fontdimen10 \scriptfont\tw@
        \advance\baselineskip by \fontdimen12 \scriptfont\tw@
        \lineskip=\thr@@\fontdimen8 \scriptfont\thr@@
        \lineskiplimit=\thr@@\fontdimen8 \scriptfont\thr@@
        \Let@\vbox\bgroup\halign\bgroup \hfil$\scriptstyle
            {##}$\hfil\cr}
\def\endSp{\crcr\egroup\egroup\egroup}
\def\Let@{\relax\iffalse{\fi\let\\=\cr\iffalse}\fi}
\def\vspace@{\def\vspace##1{\noalign{\vskip##1 }}}
\def\aligned{\,\vcenter\bgroup\vspace@\Let@\openup\jot\m@th\ialign
  \bgroup \strut\hfil$\displaystyle{##}$&$\displaystyle{{}##}$\hfil\crcr}
\def\endaligned{\crcr\egroup\egroup}
\def\matrix{\,\vcenter\bgroup\Let@\vspace@
    \normalbaselines
  \m@th\ialign\bgroup\hfil$##$\hfil&&\quad\hfil$##$\hfil\crcr
    \mathstrut\crcr\noalign{\kern-\baselineskip}}
\def\endmatrix{\crcr\mathstrut\crcr\noalign{\kern-\baselineskip}\egroup
                \egroup\,}
\newtoks\hashtoks@
\hashtoks@={#}
\def\format{\crcr\egroup\iffalse{\fi\ifnum`}=0 \fi\format@}
\def\format@#1\\{\def\preamble@{#1}%
  \def\c{\hfil$\the\hashtoks@$\hfil}%
  \def\r{\hfil$\the\hashtoks@$}%
  \def\l{$\the\hashtoks@$\hfil}%
  \setbox\z@=\hbox{\xdef\Preamble@{\preamble@}}\ifnum`{=0 \fi\iffalse}\fi
   \ialign\bgroup\span\Preamble@\crcr}

\def\cases{\left\{\,\vcenter\bgroup\vspace@
     \normalbaselines\openup\jot\m@th
       \Let@\ialign\bgroup$##$\hfil&\quad$##$\hfil\crcr
      \mathstrut\crcr\noalign{\kern-\baselineskip}}

\newif\iftagsleft@
\tagsleft@true
\def\TagsOnRight{\global\tagsleft@false}
\def\tag#1$${\iftagsleft@\leqno\else\eqno\fi
 \hbox{\def\pagebreak{\global\postdisplaypenalty-\@M}%
 \def\nopagebreak{\global\postdisplaypenalty\@M}\rm(#1\unskip)}%
  $$\postdisplaypenalty\z@\ignorespaces}
\interdisplaylinepenalty=\@M
\def\allowdisplaybreak@{\def\allowdisplaybreak{\noalign{\allowbreak}}}
\def\displaybreak@{\def\displaybreak{\noalign{\break}}}
\def\align#1\endalign{\def\tag{&}\vspace@\allowdisplaybreak@\displaybreak@
  \iftagsleft@\lalign@#1\endalign\else
   \ralign@#1\endalign\fi}
\def\ralign@#1\endalign{\displ@y\Let@\tabskip\centering\halign to\displaywidth
     {\hfil$\displaystyle{##}$\tabskip=\z@&$\displaystyle{{}##}$\hfil
       \tabskip=\centering&\llap{\hbox{(\rm##\unskip)}}\tabskip\z@\crcr
             #1\crcr}}
\def\lalign@
 #1\endalign{\displ@y\Let@\tabskip\centering\halign to \displaywidth
   {\hfil$\displaystyle{##}$\tabskip=\z@&$\displaystyle{{}##}$\hfil
   \tabskip=\centering&\kern-\displaywidth
        \rlap{\hbox{(\rm##\unskip)}}\tabskip=\displaywidth\crcr
               #1\crcr}}
\def\overrightarrow{\mathpalette\overrightarrow@}
\def\overrightarrow@#1#2{\vbox{\ialign{$##$\cr
    #1{-}\mkern-6mu\cleaders\hbox{$#1\mkern-2mu{-}\mkern-2mu$}\hfill
     \mkern-6mu{\to}\cr
     \noalign{\kern -1\p@\nointerlineskip}
     \hfil#1#2\hfil\cr}}}
\def\overleftarrow{\mathpalette\overleftarrow@}
\def\overleftarrow@#1#2{\vbox{\ialign{$##$\cr
     #1{\leftarrow}\mkern-6mu\cleaders\hbox{$#1\mkern-2mu{-}\mkern-2mu$}\hfill
      \mkern-6mu{-}\cr
     \noalign{\kern -1\p@\nointerlineskip}
     \hfil#1#2\hfil\cr}}}
\def\overleftrightarrow{\mathpalette\overleftrightarrow@}
\def\overleftrightarrow@#1#2{\vbox{\ialign{$##$\cr
     #1{\leftarrow}\mkern-6mu\cleaders\hbox{$#1\mkern-2mu{-}\mkern-2mu$}\hfill
       \mkern-6mu{\to}\cr
    \noalign{\kern -1\p@\nointerlineskip}
      \hfil#1#2\hfil\cr}}}
\def\underrightarrow{\mathpalette\underrightarrow@}
\def\underrightarrow@#1#2{\vtop{\ialign{$##$\cr
    \hfil#1#2\hfil\cr
     \noalign{\kern -1\p@\nointerlineskip}
    #1{-}\mkern-6mu\cleaders\hbox{$#1\mkern-2mu{-}\mkern-2mu$}\hfill
     \mkern-6mu{\to}\cr}}}
\def\underleftarrow{\mathpalette\underleftarrow@}
\def\underleftarrow@#1#2{\vtop{\ialign{$##$\cr
     \hfil#1#2\hfil\cr
     \noalign{\kern -1\p@\nointerlineskip}
     #1{\leftarrow}\mkern-6mu\cleaders\hbox{$#1\mkern-2mu{-}\mkern-2mu$}\hfill
      \mkern-6mu{-}\cr}}}
\def\underleftrightarrow{\mathpalette\underleftrightarrow@}
\def\underleftrightarrow@#1#2{\vtop{\ialign{$##$\cr
      \hfil#1#2\hfil\cr
    \noalign{\kern -1\p@\nointerlineskip}
     #1{\leftarrow}\mkern-6mu\cleaders\hbox{$#1\mkern-2mu{-}\mkern-2mu$}\hfill
       \mkern-6mu{\to}\cr}}}
\def\sqrt#1{\radical"270370 {#1}}
\def\dots{\relax\ifmmode\let\next=\ldots\else\let\next=\tdots@\fi\next}
\def\tdots@{\unskip\ \tdots@@}
\def\tdots@@{\futurelet\next\tdots@@@}
\def\tdots@@@{$\mathinner{\ldotp\ldotp\ldotp}\,
   \ifx\next,$\else
   \ifx\next.\,$\else
   \ifx\next;\,$\else
   \ifx\next:\,$\else
   \ifx\next?\,$\else
   \ifx\next!\,$\else
   $ \fi\fi\fi\fi\fi\fi}
\def\text{\relax\ifmmode\let\next=\text@\else\let\next=\text@@\fi\next}
\def\text@@#1{\hbox{#1}}
\def\text@#1{\mathchoice
 {\hbox{\everymath{\displaystyle}\def\textfonti{\the\textfont1 }%
    \def\textfontii{\the\textfont2 }\textdef@@ T#1}}
 {\hbox{\everymath{\textstyle}\def\textfonti{\the\textfont1 }%
    \def\textfontii{\the\textfont2 }\textdef@@ T#1}}
 {\hbox{\everymath{\scriptstyle}\def\textfonti{\the\scriptfont1 }%
   \def\textfontii{\the\scriptfont2 }\textdef@@ S\rm#1}}
 {\hbox{\everymath{\scriptscriptstyle}\def\textfonti{\the\scriptscriptfont1 }%
   \def\textfontii{\the\scriptscriptfont2 }\textdef@@ s\rm#1}}}
\def\textdef@@#1{\textdef@#1\rm \textdef@#1\bf
   \textdef@#1\sl \textdef@#1\it}

\def\textdef@#1#2{\def\next{\csname\expandafter\eat@\string#2fam\endcsname}%
\if S#1\edef#2{\the\scriptfont\next\relax}%
 \else\if s#1\edef#2{\the\scriptscriptfont\next\relax}%
 \else\edef#2{\the\textfont\next\relax}\fi\fi}
\scriptfont\itfam=\tenit \scriptscriptfont\itfam=\tenit
\scriptfont\slfam=\tensl \scriptscriptfont\slfam=\tensl
\mathcode`\0="0030
\mathcode`\1="0031
\mathcode`\2="0032
\mathcode`\3="0033
\mathcode`\4="0034
\mathcode`\5="0035
\mathcode`\6="0036
\mathcode`\7="0037
\mathcode`\8="0038
\mathcode`\9="0039
\def\Cal{\relax\ifmmode\let\next=\Cal@\else
     \def\next{\errmessage{Use \string\Cal\space only in math mode}}\fi\next}
\def\Cal@#1{{\fam2 #1}}
\def\bold{\relax\ifmmode\let\next=\bold@\else
   \def\next{\errmessage{Use \string\bold\space only in math
      mode}}\fi\next}\def\bold@#1{{\fam\bffam #1}}
\mathchardef\Gamma="0000
\mathchardef\Delta="0001
\mathchardef\Theta="0002
\mathchardef\Lambda="0003
\mathchardef\Xi="0004
\mathchardef\Pi="0005
\mathchardef\Sigma="0006
\mathchardef\Upsilon="0007
\mathchardef\Phi="0008
\mathchardef\Psi="0009
\mathchardef\Omega="000A
\mathchardef\varGamma="0100
\mathchardef\varDelta="0101
\mathchardef\varTheta="0102
\mathchardef\varLambda="0103
\mathchardef\varXi="0104
\mathchardef\varPi="0105
\mathchardef\varSigma="0106
\mathchardef\varUpsilon="0107
\mathchardef\varPhi="0108
\mathchardef\varPsi="0109
\mathchardef\varOmega="010A
\font\dummyft@=dummy
\fontdimen1 \dummyft@=\z@
\fontdimen2 \dummyft@=\z@
\fontdimen3 \dummyft@=\z@
\fontdimen4 \dummyft@=\z@
\fontdimen5 \dummyft@=\z@
\fontdimen6 \dummyft@=\z@
\fontdimen7 \dummyft@=\z@
\fontdimen8 \dummyft@=\z@
\fontdimen9 \dummyft@=\z@
\fontdimen10 \dummyft@=\z@
\fontdimen11 \dummyft@=\z@
\fontdimen12 \dummyft@=\z@
\fontdimen13 \dummyft@=\z@
\fontdimen14 \dummyft@=\z@
\fontdimen15 \dummyft@=\z@
\fontdimen16 \dummyft@=\z@
\fontdimen17 \dummyft@=\z@
\fontdimen18 \dummyft@=\z@
\fontdimen19 \dummyft@=\z@
\fontdimen20 \dummyft@=\z@
\fontdimen21 \dummyft@=\z@
\fontdimen22 \dummyft@=\z@
\def\fontlist@{\\{\tenrm}\\{\sevenrm}\\{\fiverm}\\{\teni}\\{\seveni}%
 \\{\fivei}\\{\tensy}\\{\sevensy}\\{\fivesy}\\{\tenex}\\{\tenbf}\\{\sevenbf}%
 \\{\fivebf}\\{\tensl}\\{\tenit}\\{\tensmc}}
\def\dodummy@{{\def\\##1{\global\let##1=\dummyft@}\fontlist@}}
\newif\ifsyntax@
\newcount\countxviii@
\def\newtoks@{\alloc@5\toks\toksdef\@cclvi}
\def\nopages@{\output={\setbox\z@=\box\@cclv \deadcycles=\z@}\newtoks@\output}
\def\syntax{\syntax@true\dodummy@\countxviii@=\count18
\loop \ifnum\countxviii@ > \z@ \textfont\countxviii@=\dummyft@
   \scriptfont\countxviii@=\dummyft@ \scriptscriptfont\countxviii@=\dummyft@
     \advance\countxviii@ by-\@ne\repeat
\dummyft@\tracinglostchars=\z@
  \nopages@\frenchspacing\hbadness=\@M}
\def\magstep#1{\ifcase#1 1000\or
 1200\or 1440\or 1728\or 2074\or 2488\or
 \errmessage{\string\magstep\space only works up to 5}\fi\relax}
{\lccode`\2=`\p \lccode`\3=`\t
 \lowercase{\gdef\tru@#123{#1truept}}}

\def\scaletype#1{\mag=#1\relax
 \hsize=\expandafter\tru@\the\hsize
 \vsize=\expandafter\tru@\the\vsize
 \dimen\footins=\expandafter\tru@\the\dimen\footins}

\def\scalefont#1#2\andcallit#3{\edef\font@{\the\font}#1\font#3=
  \fontname\font\space scaled #2\relax\font@}
\def\Mag@#1#2{\ifdim#1<1pt\multiply#1 #2\relax\divide#1 1000 \else
  \ifdim#1<10pt\divide#1 10 \multiply#1 #2\relax\divide#1 100\else
  \divide#1 100 \multiply#1 #2\relax\divide#1 10 \fi\fi}
\def\scalelinespacing#1{\Mag@\baselineskip{#1}\Mag@\lineskip{#1}%
  \Mag@\lineskiplimit{#1}}
\def\wlog#1{\immediate\write-1{#1}}
\catcode`\@=\active